\documentclass[11pt]{article}
\newcommand{\hf}{{1\over 2}}

\newcommand{\kim}{ k_{1}^{\mu}}                                      
\newcommand{\kom}{ k_{0}^{\mu}}                                      
\newcommand{\ki}{ k_{1}}

\newcommand{\kin}{ k_{1}^{\nu}}  
\newcommand{\kir}{ k_{1}^{\rho}} 
\newcommand{\kis}{ k_{1}^{\sigma}}                                                                                                               
 
\newcommand{\kor}{ k_{0}^{\rho}}
\newcommand{\kos}{ k_{0}^{\sigma}}                                                                                                                 
\newcommand{\ktm}{ k_{2}^{\mu}} 

\newcommand{\ktr}{ k_{2}^{\rho}} 
\newcommand{\kts}{ k_{2}^{\sigma}}

\newcommand{\p}{\partial}

\newcommand{\li}{ \lambda_{1}}                                    
\newcommand{\lt}{ \lambda_{2}}                                    
                                        
\newcommand{\al}{\alpha }

\newcommand{\be}{\begin{equation}}
\newcommand{\br}{\begin{eqnarray}}
\newcommand{\ee}{\end{equation}} 
\newcommand{\er}{\end{eqnarray}}

\begin{document}
\renewcommand{\theequation}{\thesection.\arabic{equation}}

\title{
\hfill\parbox{4cm}{\normalsize IMSC/2006/12/28\\
                               hep-th/0612069}\\        
\vspace{2cm}
 Action for (Free) Open String Modes in AdS Space Using the Loop Variable Approach.
\author{B. Sathiapalan\\ {\em Institute of Mathematical Sciences}\\
{\em Taramani}\\{\em Chennai, India 600113}\\ bala@imsc.res.in}}           
\maketitle     

\begin{abstract} 
The loop variable technique (for open strings in flat space) 
is a gauge invariant generalization of the renormalization group
method for  obtaining equations of motion. Unlike the beta functions, which are only
 proportional
to the equations of motion, here it gives the full equation of motion.   
In an earlier paper, a technique was described for adapting this method to open strings in
gravitational backgrounds.  However unlike the flat space case,
 these equations cannot be derived from an action and are therefore not complete. 
This is because there are ambiguities
in the method that involve curvature couplings that cannot be fixed by appealing to gauge
 invariance alone
but need a more complete treatment of the closed string background. An indirect
 method
to resolve these ambiguities is to require symmetricity of the second derivatives
of the action. In general this  will involve modifying the equations by terms with
 arbitrarily high powers
of curvature tensors.  This is illustrated for the massive spin 2 field. It is shown 
 that in the 
special case of an AdS or dS background, the exact action can easily be determined in
 this way.

\end{abstract}

\section{Introduction}

The loop variable method \cite{BSLV,BSREV} which is a generalization of the RG method [\cite{L} - \cite{BM}]
of obtaining gauge invariant equations of motion has been 
primarily applied to the open string in flat space. One obtains the full equation of motion rather than just the beta functions, which are only proportional to the equations.
 The method has
been generalized to closed strings. The problem of combined open and 
closed strings has not been attempted yet. However
in an earlier paper \cite{BSCurv} it was shown that a simple procedure exists for
 writing down gauge and
generally covariant equations for massive higher spin modes of the open string in 
curved spacetime.  The procedure was to first
take the loop variable equations and covariantize them in a well defined way.
 The map from loop variables to spacetime
fields has to be modified in such a way that gauge transformations can be assigned to 
spacetime fields. It was shown
that this is always possible as long as the fields are massive. This is of course the case
for the higher spin open string modes. The modifications 
involve terms that have the mass
parameter in the denominator. The zero mass limit is thus subtle 
(if at all it is well defined).
 This procedure
was illustrated in detail for a massive spin 2 field. 

There are however some ambiguities in the procedure. One is that the loop variable
 equations themselves can be modified
while preserving gauge invariance. The second is that the map to spacetime fields
 suffers from ambiguities, because in principle
there may be gauge invariant and generally covariant terms involving curvature couplings 
that can be added. These ambiguities
can in principle be fixed by modifying the loop variable equation taking into account the modified
 sigma model action - i.e. the closed string modes. Another indirect way
is to demand that the equations be derivable from an action i.e. use the fact that the
 second derivatives of the action 
(which involves first derivatives of the equations) should be symmetric in fields.
 In this paper we adopt this second procedure. In the general case
it can be seen that the procedure will involve all higher powers of the curvature tensors.
However in the special case of
symmetric spaces such as AdS or dS the procedure can be done exactly and one easily
obtains an action. Similar actions have appeared earlier in the literature \cite{Z,BKL}.

This paper is organized as follows: Section 2 is a summary of the results of \cite{BSCurv}.
 Section 3 describes 
some of the  modifications required to ensure symmetricity of the derivatives upto the point where higher powers
of the curvature appear. Section 4 completes the program for AdS (or dS) backgrounds. Section 5 has some conclusions.

\section{Review}

We give a summary of the results of \cite{BSCurv}. 

The main ingredients are the following:
\begin{enumerate}

\item
The loop variable equations for massive spin 2 in flat space read as follows: \cite{BSLV}
\br  \label{sp2}
-(k_0^2 +(k_0^5)^2)k_1^\mu k_1^\nu + k_1^{(\mu} k_0^{\nu )} k_1.k_0 - \\
 k_0^\mu k_0^\nu
k_1.k_1 + k_1^{(\mu} k_0^{\nu )} k_1^5 k_0^5 -  k_1^5 k_1^5 k_0^\mu k_0^\nu & = & 0
 \nonumber \\ \nonumber \\
k_0^2 k_2^\mu - k_1^\mu k_1.k_0 + k_0^\mu k_1.k_1 - k_0^\mu k_2.k_0  
 &=& 0  \nonumber \\
 \\
-(k_1.k_0)^2 + (k_0^2 +(k_0^5)^2)k_1.k_1 - 2k_1^5k_0^5 k_1.k_0 + 
k_0^2 k_1^5k_1^5 &=&0\nonumber \\ 
\er

Here $\langle \kim \kin \rangle = S^{\mu \nu}$, $\langle \ktm \rangle = S ^\mu$ and $\langle k_2^5 \rangle = S_5$ 
are the fields describing massive spin 2 (including the auxiliary fields.

The equations are gauge invariant under the gauge transformations:
\be
\kim \rightarrow \kim + \kom \li ~~~~\ktm \rightarrow \ktm + \kim \li + \kom \lt ~~~~~~k^5 \rightarrow k^5 + 2\lt k_0^5
\ee 

In terms of space time fields we let $\langle \li \kim \rangle = \Lambda ^\mu$
$\langle \lt \rangle = \Lambda $ and $ \kom $ is $\p ^\mu $ and $k_0^5$ 
is  equal to the mass. This gives
\be
\delta S_{\mu \nu} = \p _{(\mu} \Lambda _{\nu )} ~~~ \delta S_\mu = \p _\mu \Lambda + \Lambda _\mu ~~~~\delta S_5 = 2 \Lambda k_0^5
\ee

\item In going to curved space we let $\kom = {\p \over \p y_\mu}$ where 
$y^\mu$ is the Riemann Normal Coordinate (RNC) at some point $x_0$ which we take
as the origin of the RNC coordinate system.

We can then use the following relations to express all the terms in terms of tensors
defined at the point $x_0$:
\[
W_{\al _1 ....\al _p}(x) 
= W_{\al _1 ....\al _p}(x_0) ~+~
W_{\al _1 ....\al _p , \mu}(x_0)y^\mu ~+~
\]
\[
{1\over 2!}\{W_{\al _1 ....\al _p ,\mu \nu}(x_0) 
~-~{1\over 3}
\sum _{k=1}^p R^\beta _{~\mu \al _k \nu}(x_0) 
W_{\al _1 ..\al _{k-1}\beta \al _{k+1}..\al _p}(x_0)\}
 y^\mu y^\nu
~+~
\]
\[   
{1\over 3!}\{W_{\al _1 ....\al _p ,\mu \nu \rho}(x_0) - 
\sum _{k=1}^p R^\beta _{~\mu \al _k \nu}(x_0) 
W_{\al _1 ..\al _{k-1}\beta \al _{k+1}..\al _p, \rho}(x_0)
\]
\be \label{Taylor}
{1\over 2}\sum _{k=1}^p R^\beta _{~\mu \al _k \nu ,\rho }(x_0) 
W_{\al _1 ..\al _{k-1}\beta \al _{k+1}..\al _p}(x_0)\}y^\mu y^\nu y^\rho +...
\ee

Thus for instance it is true that $\Gamma _{\mu \nu}^\sigma |_{x_0}=0$ but
\be
\p _\rho \Gamma _{\mu \nu}^{\sigma}|_{x_0} = {1\over 3} R ^\sigma _{(\mu |\rho| \nu)}(x_0)
\ee
where we have indicated, by the curved brackets, 
symmetrization over $\mu ,\nu$ and, because of the vertical lines,  $\rho$ is excluded from symmetrization. 

This gives the following field equations:
\[
H_{\mu \nu}  \equiv 
\]
\[
-(D^\rho D_\rho + (k_0^5)^2)S_{\mu \nu} + D^\rho D_{(\mu}S_{\nu ) \rho} - {1\over 2} D_{(\mu}D_{\nu )} S^\rho _\rho
\]
\be \label{H2}
+ R^{\beta ~~ \rho}_{~ (\nu ~ \mu )} S_{\beta \rho} + D_{(\mu }S_{\nu )} (k_0^5)^2 - D_\mu D_\nu S_5 (k_0^5)  =0
\ee 

\be   \label{H1}
 H_\mu \equiv 
\]
\[
D^\rho D_\rho S_\mu - D^\rho S_{\mu \rho} + D_\mu S^\rho _\rho - 
D_\mu D^\rho S_\rho - R^\rho _\mu S_\rho  =0
\ee

\[
H
\equiv
\]
\[ 
(D^\rho D_\rho + (k_0^5)^2)  S^\sigma _\sigma - D^\rho D^\sigma S_{\rho \sigma} 
\]
\be   \label{H0}
- 2 D^\rho S_\rho (k_0^5)^2 + D^\rho D_\rho S_5 k_0^5 - R^{\alpha \beta}S_{\alpha \beta}   =0
\ee

\item
The covariantized gauge variation:  
\be  \label{GT}
\delta S_{\mu \nu}= D_{(\mu} \Lambda _{\nu )}
\ee along with corresponding
ones for the other fields, does not leave the equations invariant. This is because of the curvature
terms in (\ref{Taylor}).  The solution given in \cite{BSCurv} was to modify the map from loop variables
to space time fields. We reproduce those results (after correcting  errors in some of the expressions given there):

Thus we let  
 \be   \label{F1}
\langle \kom \kin \ki ^\rho \rangle =  D^ \mu S^{\nu \rho} + {2\over 3} R^{\lambda (\nu | \mu | \rho )} \tilde S_\lambda \equiv
F^{\mu \nu \rho}
\ee
 where $\tilde S_\lambda = S_\lambda - {D_\lambda S_5 \over 2 k_o^5}$

The extra curvature coupling ensures that the gauge transformation of the RHS (\ref{F1} (using (\ref{GT}) gives exactly
\be
\langle k_{0\rho} \li (k_{0\sigma} k_{0\alpha} + k_{1\sigma} k_{0\alpha} ) \rangle = D_\rho D_{(\sigma} \Lambda _{\alpha )}
+ {2\over 3}R^\lambda _{(\sigma |\rho | \alpha )} \Lambda _{\lambda} 
\ee
where we have used the usual covariantization of the RNC Taylor expansion.

Similarly in the case $k_ {0\rho }k_{0\sigma} k_{1\mu}k_{1\nu}$  of one adds a tensor $f_{\rho \sigma \mu \nu}$ and modifies the map in order to ensure the correct gauge transformation: 
\be
\langle k_{0\rho }k_{0\sigma} k_{1\mu}k_{1\nu} \rangle = D_\rho D_\sigma S_{\mu \nu} + {1\over 3}R^\beta _{(\mu |\rho |\sigma )}S_{\beta \nu}
+{1\over 3}R^\beta _{(\nu |\rho |\sigma )}S_{\mu \beta } + f_{\rho \sigma \mu \nu}
\ee

The tensor $f$ is:
\[ f_{\rho \sigma \mu \nu}
{2\over 3} R^\alpha _{(\nu |\rho |\mu )} D_\sigma \tilde S_\alpha +{2\over 3} R^\alpha _{(\nu |\sigma |\mu )} D_\rho \tilde S_\alpha
\]
\[ 
+[{1\over 2}  D_\rho R^\alpha _{(\nu |\sigma |\mu )} -{1\over 12}D_\rho R^\alpha _{(\mu \nu ) \sigma )}]\tilde S_\alpha
\]
\[
+[{1\over 6} D_\sigma R^\alpha _{(\nu |\rho |\mu )} - {1\over 12} D_\sigma R^\alpha _{(\mu \nu ) \rho} ] \tilde S_\alpha
\]
\[
+[{1\over 6} D_{( \mu}R^\alpha _{\nu ) \rho \sigma} - {1\over 12}( D_{\mu } R^\alpha _{\rho \nu \sigma} + D_{\nu } R^\alpha _{\rho \mu \sigma})
-{1\over 12} ( D_{\mu } R^\alpha _{\sigma \nu \rho} + D_{\nu } R^\alpha _{\sigma \mu \rho})]\tilde S_\alpha
\]

Using these results one finds in place of (\ref{H2}), (\ref{H1}), (\ref{H0}) the following
(although the tensor $f$ complicated, the combination $ - f^\rho _{\rho \mu \nu} + f^\rho _{(\mu \nu ) \rho} - {1\over 2} 
f_{(\mu \nu )~\rho}^{~~~~\rho}$ that occurs in $H_{\mu \nu}$ is actually quite simple as can be seen below)  :
\[
H_{\mu \nu}  \equiv 
\]
\[
-(D^\rho D_\rho + (k_0^5)^2)S_{\mu \nu} + D^\rho D_{(\mu}S_{\nu ) \rho} - {1\over 2} D_{(\mu}D_{\nu )} S^\rho _\rho
\]
\[
+ R^{\beta ~~ \rho}_{~ (\nu ~ \mu )} S_{\beta \rho} + D_{(\mu }S_{\nu )} (k_0^5)^2 - D_\mu D_\nu S_5 (k_0^5) 
\]
\be
-2 R^{\alpha ~~\rho}_{~(\nu ~\mu)} D_\rho \tilde S_\alpha  - D_\rho R^{\alpha ~~\rho}_{~(\nu ~\mu)}  \tilde S_\alpha 
-2 R^\alpha _{(\mu} D_{\nu )} \tilde S_\alpha  - D_{(\nu }R^\alpha _{\mu )}  \tilde S_\alpha =0.
\ee
\be   \label{H1}
 H_\mu \equiv 
\]
\[
D^\rho D_\rho S_\mu - D^\rho S_{\mu \rho} + D_\mu S^\rho _\rho - 
D_\mu D^\rho S_\rho - R^\rho _\mu S_\rho  + 2 R^\rho _\mu \tilde S _\rho  =0
\ee
\[
H
\equiv
\]
\[ 
(D^\rho D_\rho + (k_0^5)^2)  S^\sigma _\sigma - D^\rho D^\sigma S_{\rho \sigma} 
\]
\be   \label{H0}
- 2 D^\rho S_\rho (k_0^5)^2 + D^\rho D_\rho S_5 k_0^5 - R^{\alpha \beta}S_{\alpha \beta} +4 R^{\alpha \rho} D_\rho \tilde S_\alpha +
 2   D_\rho R^{\alpha \rho} \tilde S_\alpha   =0
\ee

These equations are gauge invariant, (manifestly) general coordinate covariant, and reduce to the correct equations
in the flat space limit. This concludes our summary of \cite{BSCurv}.

\end{enumerate}
\section{Towards an Action}

The question we can now ask is whether these equations can be obtained from an action. This is equivalent to
checking whether ${\p H_j \over \p \phi _i} = {\p  H_i \over \p \phi _j} $. If $H_i = {\p  S \over \p \phi _i}$, this 
would be true.  It is easy to see that these equations do not satisfy these conditions. In fact even the flat space
equations need to be redefined for this to be true. Thus we define 
\be
H'_{\mu \nu} = H_{\mu \nu } + g_{\mu \nu} H ~~~~ H' = H + g^{\mu \nu} H_{\mu \nu}
\ee
Thus we get:
\[
H'_{\mu \nu}  \equiv 
\]
\[
-(D^\rho D_\rho + (k_0^5)^2)S_{\mu \nu} + D^\rho D_{(\mu}S_{\nu ) \rho} - {1\over 2} D_{(\mu}D_{\nu )} S^\rho _\rho
\]
\[
+ R^{\beta ~~ \rho}_{~ (\nu ~ \mu )} S_{\beta \rho} + D_{(\mu }S_{\nu )} (k_0^5)^2 - D_\mu D_\nu S_5 (k_0^5) 
\]
\[
-2 R^{\alpha ~~\rho}_{~(\nu ~\mu)} D_\rho \tilde S_\alpha  - D_\rho R^{\alpha ~~\rho}_{~(\nu ~\mu)}  \tilde S_\alpha 
-2 R^\alpha _{(\mu} D_{\nu )} \tilde S_\alpha  - D_{(\nu }R^\alpha _{\mu )}  \tilde S_\alpha 
\]
\[ 
+g_{\mu \nu}\Big [(D^\rho D_\rho + (k_0^5)^2)  S^\sigma _\sigma - D^\rho D^\sigma S_{\rho \sigma} 
\]
\be   \label{H0}
- 2 D^\rho S_\rho (k_0^5)^2 + D^\rho D_\rho S_5 k_0^5 - R^{\alpha \beta}S_{\alpha \beta} +4 R^{\alpha \rho} D_\rho \tilde S_\alpha +
 2   D_\rho R^{\alpha \rho} \tilde S_\alpha \Big ]  =0
\ee
$H'$ actually simplifies to:
\[
H' \equiv
\]
\be
D^\rho D^\sigma S_{\rho \sigma} - D^\rho D_\rho S^\sigma _\sigma + R^{\alpha \beta} S_{\alpha \beta} -
 4 R^{\alpha \rho}D_\rho \tilde S_{\alpha }-2 D^\rho R^\alpha _\rho \tilde S_\alpha =0
\ee
The flat space limit of these equations can be derived from an action
(actually $H'$ needs to be scaled by a factor of $k_0^5\over 2$ , but we will deal with this in section 4.).

In curved spacetime one can check that 
\[
{\p H'_{\mu \nu} (x) \over \p S^\alpha (y)} = g_{\alpha (\mu}D_{\nu )} \delta (x-y) (k_0^5)^2 -2 (k_0^5)^2 g_{\mu \nu} D_\alpha
\delta (x-y) 
\]
\[
-2R^{\alpha ~~\rho}_{(\nu ~ \mu )} D_{\rho} \delta (x-y) - D_\rho R^{\alpha ~~\rho}_{(\nu ~ \mu )}\delta (x-y)
\]
\be  \label{I}
-2 R^\alpha _{~(\mu } D_{\alpha )} \delta (x-y) - D_{(\mu}R^\alpha _{~~~\nu )}\delta(x-y)
+g_{\mu \nu} (4 R_\alpha ^\rho D_\rho + 2 D^\rho R_{\alpha \rho} ) \delta (x-y)
\ee
which is not equal to 
\be  \label{II}
{\p H_\alpha (y) \over \p S^{\mu \nu}(x)} = -[ g_{\alpha (\mu}D_{\nu )} \delta (x-y) (k_0^5)^2 -2 (k_0^5)^2 g_{\mu \nu} D_\alpha
\delta (x-y)] 
\ee
because of the curvature couplings (overall sign can be taken care of easily).

At this point we should point out that there are two ambiguities in the procedure outlined above. One is that
the loop variable equation (2.03) for the vector can actually be generalized to:
\be
\kor \kos \ktm - {1\over 2}(\kir \kos + \kis \kor ) + \kir \kis \kom - {1\over 2} (\ktr \kos + \kts \kor ) \kom =0
\ee

Thus instead of contracting with $g^{\rho \sigma}$ one can use more generally $X^{\alpha \rho \sigma \mu}$. 
The second ambiguity is that the combination $S_{\mu \nu} - D_{(\mu} \tilde S_{\nu )}$ is gauge invariant and one can
add terms involving this to the equations of motion. Thus if there had been a term in the action
of the form 
\[
{1\over 2} (S_{\alpha \rho} - D_{(\rho } \tilde S_{\alpha )} ) T^{\alpha \nu \rho \mu}(S_{\mu \nu} - D_{(\mu} \tilde S_{\nu )})
\]
this would contribute a term 
\be   \label{T4}
(D_\rho T^{\alpha \nu \rho \mu})(S_{\mu \nu} - D_{(\mu} \tilde S_{\nu )}) + 
T^{\alpha \nu \rho \mu}D_\rho (S_{\mu \nu} - D_{(\mu} \tilde S_{\nu )})
\ee
to $H_\alpha$.

Similarly 
\be 
{1\over 2} (S_{\alpha \mu} - D_{(\mu } \tilde S_{\alpha )} ) T^{\alpha  \mu}(S_{\mu \nu} - D_{(\mu} \tilde S_{\nu )})
\ee
contributes a term
\be  \label{T2}
(D_{(\mu} T^{\alpha }_{\nu )})(S_{\mu \nu} - D_{(\mu} \tilde S_{\nu )}) + 
T^{\alpha }_{(\nu }D_{\mu)} (S_{\mu \nu} - D_{(\mu} \tilde S_{\nu )})
\ee
to $H_\alpha$.

One can see that these are the right kinds of terms that one needs to get agreement between (\ref{I}) and (\ref{II}).
The coefficients of $D_\rho R S $ and $R D_\rho S$ are not equal (in (\ref{I}), therefore (\ref{T4}) and (\ref{T2})
by themselves cannot be sufficient - we will also need terms involving $R$ in the coefficient tensor $X^{\alpha \rho \sigma \mu}$
that multiplies the vector equation. 

If we assume that $X$ and $T$ are linear in the curvature tensor then one can write the most general
form for $X$ and $T$ and attempt to satisfy the integrability condition.\footnote{One should also ensure that the symmetry properties
of the tensor $T$ are such that terms involving more than two derivatives of $S_5$ actually vanish.} 

The result of doing this is that if one chooses 
\be
X^{\alpha \nu \rho \mu} = -(k_0^5)^2 g^{\nu \rho} g^{\alpha \mu} + {2\over 3} R^{\alpha \nu \rho \mu} + g^{\nu \rho} R^{\alpha \mu}
\ee
\be
T^\alpha _{(\mu |\rho | \nu )} = - R^\alpha _{(\mu |\rho |\nu )} + 2 g_{\mu \nu} R_{\alpha \rho}
~~;~~~~T^{\alpha \nu} = - {1\over 2} R^{\alpha \nu} 
\ee 
one finds that 
\[
{\p H_\alpha (y) \over \p S^{\mu \nu}(x)}={\p H'_{\mu \nu} (x) \over \p S^\alpha (y)}
\]

With this our equation for the vector becomes:
\[ 
H_\alpha \equiv 
\]
\[
X^{\alpha \nu \rho \beta} \Big [ {1\over 2} D_{(\rho }D_{\sigma )} S_\beta - {1\over 2} D_{(\rho} S_{\sigma ) \beta} +
D_\beta S_{\rho \sigma} - {1\over 2} D_\beta D_{(\rho }S_{\sigma )} 
\]
\[
+ {1\over 2} R^\lambda _{(\sigma \rho ) \beta} S_{\lambda} - R^\lambda _{(\sigma \rho ) \beta} \tilde S _\lambda \Big ]
\]
\[
(D_\rho T^{\alpha \nu \rho \mu})(S_{\mu \nu} - D_{(\mu} \tilde S_{\nu )}) + 
T^{\alpha \nu \rho \mu}D_\rho (S_{\mu \nu} - D_{(\mu} \tilde S_{\nu )})
\]
\be  \label{vec}
(D_{(\mu} T^{\alpha }_{\nu )})(S_{\mu \nu} - D_{(\mu} \tilde S_{\nu )}) + 
T^{\alpha }_{(\nu }D_{\mu)} (S_{\mu \nu} - D_{(\mu} \tilde S_{\nu )}) =0
\ee

One immediately sees that $R^2$ terms have appeared in the equation. In particular the coupling to the scalar $S_5$
(contained in $\tilde S _\lambda$) has an $R^2$ piece. The scalar equation, which was linear in the  curvature tensor,
will now have to be modified. This will have to involve the gauge invariant combination of fields 
$S_{\mu \nu} - D_{(\mu} \tilde S_{\nu )}$. This would then modify the coupling to $S_{\mu \nu}$ by $R^2$ terms.
In this manner one can presumably modify the equations iteratively, in powers of $R$, and one expects some non polynomial
(in $R$) result. We will not attempt to do this in this paper. In the next section we shall specialize to AdS space where we will
see that the process terminates in one iteration and a gauge invariant action can be written.

\section{AdS Action}

In AdS space we can take \footnote{By changing the sign of $L^2$ these results apply also to de Sitter space.}
\[
R_{\mu \nu \rho \sigma} = -{1\over L^2} (g_{\mu \rho} g_{\nu \sigma } - g_{\mu \sigma} g_{\nu \rho}) 
\]
\be
R_{\mu \rho} = g^{\nu \sigma } R_{\mu \nu \rho \sigma} = -{(D-1)\over L^2} g_{\mu \rho}
\ee

We can use this form in the tensor $X$ which becomes ($b=(D-1)$):
\be
X^{\alpha \rho \sigma \beta} = -\big [ {b-2/3\over L^2} + (k_0^5)^2 \big ] g^{\rho \sigma}g^{\alpha \beta} -
{2\over 3L^2} g^{\alpha \sigma}g^{\rho \beta}
\ee
The contribution to the equation of motion $H_\alpha$ due to the X - term can be worked out. It is:

\be
-[{b-1\over L^2} + (k_0^5)^2] \big [ D^\rho D_\rho S_\mu - D^\rho S_{\alpha \rho} + D_\alpha S^\rho _\rho - 
D_\alpha D^\rho S_\rho - {2b\over L^2} ({S_\alpha \over 2} - \tilde S_\alpha)\big ]  =0
\ee

Similarly the contribution from the terms involving $T$ turn out to have the same structure:

\be
{(1-b)\over L^2} \big [ D^\rho D_\rho S_\mu - D^\rho S_{\alpha \rho} + D_\alpha S^\rho _\rho - 
D_\alpha D^\rho S_\rho - {2b\over L^2} ({S_\alpha \over 2} - \tilde S_\alpha)\big ]  =0
\ee

Adding the two then gives $H_\alpha$:
\[
H_\alpha \equiv
\]
\be
 -[{2(b-1)\over L^2} + (k_0^5)^2] \big [ D^\rho D_\rho S_\alpha - D^\rho S_{\alpha \rho} + D_\alpha S^\rho _\rho - 
D_\alpha D^\rho S_\rho - {2b\over L^2} ({S_\alpha \over 2} - \tilde S_\alpha)\big ]  =0
\ee

Since this equation is already compatible with $H'_{\mu \nu}$ we turn to the scalar equation $H'$
specialized to AdS:

\[
H' \equiv
\]
\be   \label{h'}
D^\rho D^\sigma S_{\rho \sigma} - D^\rho D_\rho S^\sigma _\sigma  -{b\over L^2}  S^\sigma _\sigma +
 4{b\over L^2} D^\rho \tilde S _\rho  =0
\ee

Similarly the tensor equation is:
\[
H'_{\mu \nu}  \equiv 
\]
\[
-(D^\rho D_\rho + (k_0^5)^2)S_{\mu \nu} + D^\rho D_{(\mu}S_{\nu ) \rho} - {1\over 2} D_{(\mu}D_{\nu )} S^\rho _\rho
\]
\[
-{2\over L^2}(g_{\mu \nu} S^\rho _\rho - S _{\mu \nu} )
\]
\[
   {2\over L^2}[2g_{\mu \nu} D^\rho \tilde S_\rho]  + 
{(2(b-1)+ (k_0^5)^2)\over L^2}D_{(\nu }\tilde S_{\mu )}
\]
\[ 
+g_{\mu \nu}\Big [(D^\rho D_\rho + (k_0^5)^2 + {b\over L^2}))  S^\sigma _\sigma - D^\rho D^\sigma S_{\rho \sigma} 
\]
\[   
-{4b\over L^2} D^\rho \tilde S_\rho
-2D^\rho S_\rho (k_0^5)^2 + D^\rho D_\rho S_5 k_0^5 \Big ]
\]

\[=
-(D^\rho D_\rho + (k_0^5)^2)S_{\mu \nu} + D^\rho D_{(\mu}S_{\nu ) \rho} - {1\over 2} D_{(\mu}D_{\nu )} S^\rho _\rho
\]
\[
-{2\over L^2}(g_{\mu \nu} S^\rho _\rho - S _{\mu \nu} )
\]
\[   
+({2(b-1) \over L^2}+ (k_0^5)^2)D_{(\nu }\tilde S_{\mu )}
\]
\[ 
+g_{\mu \nu}\Big [(D^\rho D_\rho + (k_0^5)^2 + {b\over L^2}))  S^\sigma _\sigma - D^\rho D^\sigma S_{\rho \sigma} 
\]
\be   \label{H0}
-2({2(b-1) \over L^2}+(k_0^5)^2) D^\rho \tilde S_\rho
\Big ]=0
\ee
{\bf Compatibility of Scalar and Tensor equations:}

Again If one calculates 
\be
{\p H' \over \p S^{\mu \nu}} = k_0^5[D_{(\mu} D_{\nu)} - 2g_{\mu \nu} D^\rho D_\rho ]\delta (x-y) -{2b\over L^2} k_0^5 g_{\mu \nu}\delta (x-y)
\ee 

Note that we have multiplied the scalar equation (\ref{h'}) by $k_0^5$.
 The last term in the above expression is missing in ${\p H'_{\mu \nu} \over \p S_5}$ as one can easily check. Thus one has to modify
these two equations. There are two obvious modifications that are possible.
 One is  a field redefinition: $S_{\mu \nu} = S'_{\mu \nu} + x g_{\mu \nu} S_5$.  Another is to modify the $H'$ equation by terms that
could potentially come from  terms of the form (\ref{T4}), (\ref{T2}). They contribute linear combinations:

\[
u D^\rho D_\rho [S^\sigma _\sigma - 2 D^\sigma \tilde S_\sigma ] + v [ D^\rho D^\sigma - D^\rho D^\sigma D_{(\rho }\tilde S_{\sigma )}
\]

If one chooses $u+v=0$ the higher derivative terms cancel. Thus we get the contribution.
\[
u[D^\rho D^\rho S^\sigma _\sigma k_0^5 - k_0^5 D^\rho D ^\sigma S_{\rho \sigma} ] -2u{b\over L^2} D^\rho \tilde S_\rho k_0^5
\]

Thus we make both changes : Add the above term to the scalar equation, and replace $S_{\mu \nu}$ by $S'_{\mu \nu} + x g_{\mu \nu} S_5$
in all the equations.

We then get:
\be
{\p H' \over \p S^{\mu \nu}} = k_0^5 (1-u) [D_{(\mu} D_{\nu)} - 2g_{\mu \nu} D^\rho D_\rho ]\delta (x-y) 
 -{2b\over L^2} k_0^5 g_{\mu \nu}\delta (x-y)
\ee 
\[
{\p H'_{\mu \nu} \over \p S_5}= ({x(1-b)\over 2} - {k_0^5 \over 2} + {2(1-b)\over L^2 2k_0^5})
  [D_{(\mu} D_{\nu)} - 2g_{\mu \nu} D^\rho D_\rho ]\delta (x-y)
\]
\be
 +bxg_{\mu \nu} [ (k_0^5)^2 + {(b-1)\over L^2}]\delta (x-y)
\ee

Requiring that the corresponding coefficients be proportional (they don't have to be equal
because the scalar equation can be scaled by an overall factor)  gives the following relation :

\be  \label{ST}
{1\over L^2 k_0^5}[(k_0^5)^2 + {2(b-1)\over L^2}] = x[(k_0^5)^2 (1-u) - {u(b-1) \over L^2}]
\ee

{\bf Compatibility of Scalar and Vector Equations:}

\[
{\p H_\alpha \over \p S_5} = -b(x+{1\over L^2 k_0^5} ) [ (k_0^5)^2 + {2(b-1) \over L^2}] D_\alpha \delta (x-y)
\]
\[
{\p H' \over \p S^\alpha} = {2bk_0^5\over L^2}(2-u)D_\alpha \delta(x-y)
\]

Requiring equality of these coefficients one gets
\be   \label{SV}
{2bk_0^5\over L^2}(2-u) =-b(x+{1\over L^2 k_0^5} ) [ (k_0^5)^2 + {2(b-1) \over L^2}]
\ee

Solving (\ref{ST}) and (\ref{SV}) gives

\be
\bf  u=2  ~~~;~~~x=-{1\over k_0^5 L^2}
\ee.

Plugging in the rescaling factor, one also concludes that the actual scalar equation is: 
\be
{\delta S\over \delta S_5} = {1\over 2(k_0^5)^2}[(k_0^5)^2 + {(b-1)\over L^2}] H'
\ee
 Note also that in the final solution the vector and scalar are decoupled exactly as in the flat space limit.

After making these changes the results are as follows:

{\bf Tensor Equation:}
\[H'_{\mu \nu}\equiv\]

\[
-(D^\rho D_\rho + (k_0^5)^2)S_{\mu \nu} + D^\rho D_{(\mu}S_{\nu ) \rho} - {1\over 2} D_{(\mu}D_{\nu )} S^\rho _\rho -g_{\mu \nu}D^\rho D^\sigma S_{\rho \sigma}
\]
\[
-{2\over L^2}(g_{\mu \nu} S^\rho _\rho - S _{\mu \nu} )
\]
\[   
+({2(D-2) \over L^2}+ (k_0^5)^2)[D_{(\nu } S_{\mu )} -2g_{\mu \nu} D^\rho  S_\rho ]
\]
\[ 
+g_{\mu \nu}\Big [(D^\rho D_\rho + (k_0^5)^2 + {D-1\over L^2}))  S^\sigma _\sigma  \Big ]
\]
\be   
 + {1\over k_0^5}[{(D-2)\over L^2} + (k_0^5)^2][g_{\mu \nu} D^\rho D_\rho S_5
- D_\mu D_\nu S_5 - {D-1\over L^2} g_{\mu \nu } S_5 ]
=0
\ee

{\bf Vector Equation:}

\[
H_\alpha \equiv
\]
\be
 -[{2(D-2)\over L^2} + (k_0^5)^2] \big [ D^\rho D_\rho S_\alpha - D^\rho S_{\alpha \rho} + D_\alpha S^\rho _\rho - 
D_\alpha D^\rho S_\rho - {D-1\over L^2} S_\alpha \big ]  =0
\ee

{\bf Scalar Equation:}

\[
H' \equiv
\]
\be   \label{h'}
{1\over 2k_0^5}({(D-2)\over L^2}+ (k_0^5)^2)\big [
-D^\rho D^\sigma S_{\rho \sigma} + D^\rho D_\rho S^\sigma _\sigma  -{D-1\over L^2}  S^\sigma _\sigma \big ]
   =0
\ee

{\bf Action:}

\[ S = \int d^Dx \Big \{ -\hf S^{\mu \nu} [D^\rho D_\rho + (k_0^5)^2] S_{\mu \nu} + \hf S^{\mu \nu}D^\rho D_\mu
S_{\nu \rho} - \hf S^{\alpha \beta}D_\alpha D_\beta S^\sigma_\sigma \]
\[ + {1\over 2L^2} (S^{\rho \sigma}S_{\rho \sigma}- S^\sigma _\sigma S^\rho _\rho )+({2(D-2)\over L^2} +(k_0^5)^2)
(S^{\alpha \beta}D_\alpha S_\beta - S^\sigma _\sigma D^\rho S_\rho ) \]
\[ + {1\over 4}
S^\alpha _\alpha (D^\rho D_\rho + (k_0^5)^2 + {D-1\over L^2})S^\rho _\rho \]
\[ +{1\over 2k_0^5}({(D-2)\over L^2} +(k_0^5)^2)
[ S^\sigma _\sigma D^\rho D_\rho S_5 - S^{\rho \sigma } D_\rho D_\sigma S_5  - {D-1\over L^2} S^\sigma _\sigma S_5 ]
\]
\be
({2(D-2)\over L^2} + (k_0^5)^2)\big [-\hf D_\rho S_\sigma D^\rho S^\sigma +
 \hf D_\alpha S^\alpha D^\rho S_\rho + {D-1\over 2L^2}S_\alpha S^\alpha \big ] \Big \}
\ee 

Note that the gauge transformation of the tensor is different after the field redefinition:
\[ 
\delta S_{\mu \nu} = \p _{(\mu }\Lambda _{\nu )} + {2\over L^2} g_{\mu \nu}\Lambda
\]

\section{Conclusions}

As explained in the introduction, in flat space the loop variable method for open strings
 gives the
full equation of motion, unlike the beta function, which is only proportional to it.
 So one can extract an action from this information. In curved space
things are more complicated. One has to include the bulk two dimensional field theory
operators in addition to the boundary terms.
 In principle one should be able to get the equations by
including the RG action on the combined closed-open two dimensional field theory. 
This seems quite involved. 
In an earlier paper \cite{BSCurv}  we described a simple procedure to generalize the flat space equations
 to curved space equations. The main idea was to interpret the loop variable
momenta as being conjugate to Riemann Normal Coordinates. It was also shown there
that it is possible to maintain gauge invariance in curved space - which is usually the
stumbling block. However the question of an action was not addressed.

In this paper we have addressed this question. The equations obtained earlier (in \cite{BSCurv})
cannot be obtained from an action. They have to be modified. We have shown that this is possible 
because there are in fact  ambiguities in the procedure of \cite{BSCurv} that can be exploited. 
One can modify the equations by gauge invariant terms
that involve cuvature tensors. One can fix these ambiguities by imposing
 the requirement of compatibility of the equations, i.e by the requirement that
the equations be derivable from an action. This implies that the second derivatives
have to be symmetric. We discussed the spin two case for general backgrounds 
and showed that this leads to inclusion of terms involving
 curvature. It is an iterative process and for general backgrounds involves
arbitrary high powers of the curvature tensors.
 However in the AdS case the method gives an exact form of the action and we recover 
known results quite easily. 

This leads to many open questions. One question is whether for general backgrounds
one can obtain all the correction terms. More interestingly
the loop variable method can be easily generalized to the interacting string
 (in curved spacetime) case. 
The problem of obtaining an action has to be
addressed there also.  Since we have an infinite number of coupled equations it is 
likely that this procedure
of requiring that the equations be derivable from an action would be impractical.
 One should perhaps go back to the ``first principles''
calculation of the gauge invariant exact RG equations of the two dimensional field theory.
One should also keep in mind that in string theory interaction with gravity implies
that the string is self interacting since both interactions are generically governed
by the same coupling constant. 

{\bf Acknowledgements:} We acknowledge useful discussions with S. Kalyana Rama.

\end{document}